\documentclass[final,authoryear,5p]{elsarticle}

\usepackage{graphicx}

\usepackage{amssymb}
\usepackage{amsmath}
\usepackage{pst-eps}

\usepackage[ps2pdf,%
a4paper=true,%
breaklinks=true,%
colorlinks=true,%
pdfauthor={First Author et al.},%
pdftitle={Template for manuscripts in Advances in Space Research}%
]{hyperref}

\journal{Advances in Space Research}

\newcommand{\de}{\mathrm{d}}
\newcommand{\zav}[1]{\left(#1\right)}
\newcommand{\hzav}[1]{\left[#1\right]}
\newcommand{\varv}{v}
\newcommand\x[1]{\ensuremath{#1_\text{X}}}
\newcommand\lx{\ensuremath{\x L}}

\DeclareMathAlphabet{\mathsc}{OT1}{cmr}{m}{sc}
\def\testbx{bx}%
\DeclareRobustCommand{\ion}[2]{%
\relax\ifmmode
\ifx\testbx\f@series
{\mathbf{#1\,\mathsc{#2}}}\else
{\mathrm{#1\,\mathsc{#2}}}\fi
\else{#1\,{\scshape{#2}}}%
\fi}
\begin{document}

%%%%%%%%%%%%%%%%%%%%%%%%%%%%%%%%%%%%%%%%%%%%%%%%%%%%%%%%%%%%%%%%%%%%%%%%%%%%%
%% Frontmatter
\begin{frontmatter}


% Use the tnoteref command within \title and fnref within \author or \address for footnotes;
% use the corref command within \author for corresponding author footnotes;
% \author{Name\corref{cor1}\fnref{label2}}
% \address{Address\fnref{label3}}

\title{Influence of X-ray radiation on the
%Kr1:
hot star
wind ionization state and on the
radiative force}

% Use optional labels to link authors explicitly to addresses:

\author{Ji\v r\'\i\ Krti\v cka\corref{cor}\fnref{footnote2}}
\address{\'Ustav teoretick\'e fyziky a astrofyziky, Masarykova univerzita,
           Kotl\'a\v rsk\' a 2, CZ-611\,37 Brno, Czech
                      Republic}
\cortext[cor]{Corresponding author}
\ead{krticka@physics.muni.cz}

% Url can be given like this:
% \ead[url]{http://www.elsevier.com/wps/find/authorsview.authors/latex}

\author{Ji\v r\'\i\ Kub\'at}
\address{Astronomick\'y \'ustav, Akademie v\v{e}d \v{C}esk\'e
           republiky, Fri\v{c}ova 298, CZ-251 65 Ond\v{r}ejov, Czech Republic}
\ead{kubat@sunstel.asu.cas.cz}

\begin{abstract}
Hot stars emit large amounts of X-rays, which are assumed to originate in the
supersonic stellar wind. Part of the emitted X-rays is subsequently absorbed in
the wind and influences its ionization state. Because hot star winds are driven
radiatively, the modified ionization equilibrium affects the radiative force. We
review the recent progress in modelling the influence of X-rays on the radiative
equilibrium and on the radiative force. We focus particularly on single stars
with X-rays produced in wind shocks and on binaries with massive components,
which belong to the most luminous objects in X-rays.

\end{abstract}


\end{frontmatter}

\parindent=0.5 cm

%%%%%%%%%%%%%%%%%%%%%%%%%%%%%%%%%%%%%%%%%%%%%%%%%%%%%%%%%%%%%%%%%%%%%%%%%%%%%
%% Main text
\section{Introduction}

The idea of influence of X-rays on the ionization structure of hot star winds
dates back to the beginning of modern hot star wind studies. The observations
with the {\em Copernicus} satellite revealed the problem of 
%Kr1: ``superionization''
``super-ionization''
connected with the presence of strong P~Cygni line profiles of surprisingly high
ionization species (such us \ion{N}{v} or \ion{O}{vi}) in the spectra of
early-type stars \citep{roger,lamor,chumeli}. The existence of these ions in the
wind was subsequently proposed to be a result of radiative ionization due to the
UV flux \citep{castion} and X-ray photoionization \citep{olson,nechumeli}.

Since then the problem of the influence of X-rays on the wind ionization
structure has been addressed also by the solution of kinetic equilibrium (NLTE)
equations. \citet{pasam} showed that the ionization fraction of \ion{N}{v}
derived from observation can be reproduced using NLTE models, but the models
with only atmospheric irradiation can not fully explain the ionization fraction
of \ion{O}{vi}. Ionization equilibrium calculations that included also the X-ray
irradiation \citep[albeit in a simplified form,][]{mekac,pakupa} predicted
\ion{O}{vi} ionization fractions and wind line profiles that agree with
observations much better. Their calculations also show that the ionization
fraction of dominant ionization states is not affected by X-ray irradiation.
Consequently, the radiative driving may proceed nearly unaffected in the
presence of X-ray irradiation.

The numerical simulations of instabilities connected with radiative driving
\citep{felpulpal} provided more reliable predictions for the X-ray irradiation,
especially for the spectral energy distribution of emitted X-rays. More
realistic prescriptions for the X-ray irradiation are used in recent NLTE wind
models \citep[e.g.,][]{pahole,krfeloskuh}, although empirically-based formulae
for the X-ray irradiation also give ionization fractions and spectra that agree
with observations \citep[e.g.,][]{hilkupa, bouhil}.

Because the X-ray radiation affects the wind ionization state, it is able to
destroy ions responsible for the wind acceleration if the X-ray source is
sufficiently strong. This happens in high-mass X-ray binaries, where the X-rays
originate in the accretion of matter on the compact component. From earlier
models \citep{natural,hacek} of the stellar wind ionization structure in
high-mass X-ray binaries \citet{ff} realized that the X-ray ionization affects
also the radiative force and provided a general picture of the structure of
circumstellar environment in such objects. Numerical simulations
\citep{blondyn,felabon} revealed a complex structure of the flow influenced by
the gravity of the compact object (accretion wake) and X-rays (photoionization
wake).

The line force in X-ray irradiated stellar winds has to be obtained from the
solution of 
%Kr1:
the
3D radiative transfer equation assuming NLTE. Because this requires
formidable effort, the realistic calculations of the line force concentrated on
the 1D problem \citep{kalmek}. While \citet{stk} provide modification for usual
force multipliers \citep[originally introduced by][and \citealt{abbpar}]{cak} in
the presence of
%Kr1:
a
strong X-ray field, \citet{mytri} showed by detailed calculation
without using force multipliers that the X-rays produced in wind shocks do not
affect the line force significantly. On the other hand, in X-ray binaries with
large X-ray luminosities the influence of the X-ray emission may lead to the
decrease of the radiative force and the 
%Kr1: wind inhibition
inhibition of the stellar wind
\citep[e.g.,][]{dvojvit}.

Some attention was also given to the importance of XUV/EUV radiation for the
wind ionization balance following the work of \citet{pakupa}. The XUV/EUV
radiation was studied in connection with the ultraviolet resonance doublet of
\ion{P}{v}, which is too weak in comparison with the theory \citep{fullmap}.
\citet{walc} proposed that XUV radiation, which cannot be directly observed, may
explain the observations of \ion{P}{v} lines. However, subsequent calculations
\citep{bouhil,fosfor} showed that this explanation is unlikely, because the
XUV/EUV radiation in 
%Kr1:? ... on the model of ... in an amount ...
%Ku2: Asi myslel odkaz na model, podle ktereho to tak je.
an
amount that explains the weak observed \ion{P}{v} lines
destroys ions responsible for the wind acceleration.

In this review we focus on understanding how the high energy (X-ray and EUV)
radiation affects the ionization equilibrium and the radiative force. For
concreteness we will focus on NLTE models with comoving-frame line force of
\citet{fosfor}, which use X-ray irradiation based on model of \citet{felpulpal}
or assume
%Kr1:
a
power-law external irradiation. However, corresponding results for
ionization structure can be derived also with other codes.

%=======================================================================
\section{\ion{N}{v} ionization fraction and no need for additional
ionizing radiation sources}
\label{kadiskat}

Historically, the observations of lines of ions with higher degree of ionization
(e.g, \ion{C}{v}, \ion{N}{v}) were used as an argument for the existence of
%Kr1
an
additional source of ionizing radiation. However, this argument is based just on
an oversimplification of kinetic (NLTE) equations, which can be avoided only by
a detailed numerical analysis.

We demonstrate this on the ionization ratio of \ion{N}{iv} and \ion{N}{v}. Let
us assume that the populations of the ground levels dominate for these ions,
i.e. populations of excited states can be neglected. Let us also assume that the
collisional rates can be neglected. Both these assumptions are justified in
stellar winds. In such case the ionization balance between the \ion{N}{iv} and
\ion{N}{v} ions follows from the kinetic equilibrium equations
\citep{hubenymihalas} as
\begin{equation}
\label{nlte45}
N_4 R_{45}-N_5R_{54}=0,
\end{equation}
where $N_4$ and $N_5$ are number densities of \ion{N}{iv} and \ion{N}{v} ions,
respectively,
\begin{equation}
\label{zarion}
R_{45} = 4\pi \int_{\nu_4}^{\infty} \frac{\alpha_{4}(\nu)}{h\nu}
J(\nu)\,\de\nu
\end{equation}
is the radiative ionization rate, and
\begin{equation}
\label{zarek}
R_{54} =4\pi\! {\zav{\!\frac{N_4}{N_5}\!}\!\!}^\ast\!
\int_{\nu_4}^{\infty}\! \frac{\alpha_4(\nu)}{h\nu}\hzav{\frac{2h\nu^3}{c^2}\!+\!
J(\nu)}\! e^{-\frac{h\nu}{kT}}
\,\de\nu
\end{equation}
is the radiative recombination rate (an asterisk denotes the LTE value). In
these expressions, $\alpha_{4}(\nu)$ is the photoionization cross-section with
the photoionization edge at the frequency $\nu_4$, and $J(\nu)$ is the mean
intensity of radiation. Approximating the integrals in Eqs.~\eqref{zarion} and
\eqref{zarek} by values of the integrands at $\nu_4$ and taking into account
that for this frequency (high radiation energy) holds ${2h\nu^3}/{c^2}\gg
J(\nu)$, we derive from Eq.~\eqref{nlte45}
\begin{equation}
\frac{N_5}{N_4}= J(\nu_4) \frac{c^2}{2h\nu_4^3}{\zav{\frac{N_5}{N_4}}\!}^\ast
e^{\frac{h\nu_4}{kT}}.
\end{equation}
The fraction $({N_5}/{N_4})^*$ can be evaluated using the Saha equation
$({N_5}/{N_4})^*=2\zav{2\pi m_\text{e}kT/h^2}^{3/2}/N_\text{e}
\exp\zav{-\frac{h\nu_4}{kT}},$
yielding
\begin{equation}
\label{neelix}
\frac{N_5}{N_4}=\frac{c^2}{h^4\nu_4^3}\zav{2\pi m_\text{e}kT}^{3/2}
\frac{J(\nu_4)}{N_\text{e}},
\end{equation}
where we assumed unity ionic partition functions and $N_\text{e}$ is the free
electron number density. From this equation it seems that the ionization ratio
is directly proportional to the mean radiation intensity $J$ at a given
ionization frequency $\nu_4$. Using values appropriate for the model supergiant
with $T_\text{eff}=40\,000\,\text{K}$ \citep[with stellar mass $M_\ast$ and
radius $R_\ast$ from][]{okali} at the radius of roughly $1.1\,R_*$, namely
$\nu_4=1.9\times10^{16}\,\text{Hz}$,
$T=33\,000\,\text{K}$,
$N_\text{e}=2\times10^{11}\,\text{cm}^{-3}$, and $J(\nu_4)
=5\times10^{-12}\,\text{erg}\,\text{cm}^{-2}\,\text{s}^{-1}\,\text{Hz}^{-1}$
we obtain in absence of additional sources of ionizing radiation from
Eq.~\eqref{neelix} the ionization ratio ${N_5}/{N_4}\approx 7 \times 10^{-3}$.
This indicates low \ion{N}{v} ionization fraction and a need for higher ionizing
radiation to enhance the \ion{N}{v} abundance to values inferred from
observations. However, from our full NLTE models calculated using our code
\citep{cmf1}, which consider reliable model ions \citep[from][]{ostar2002} with
%Kr1:
a
sufficient number of energy levels, we obtain for the same wind location
${N_5}/{N_4}\approx0.14$.

The reason for the difference is that the kinetic equilibrium equations are
quite complex and their oversimplification may lead to incorrect results. In the
particular case of \ion{N}{v} ionization fraction, the most important ionization
process is not that from the ground level, but those from excited levels.
Although upper levels are less populated, they are closely coupled with the
ground level by strong bound-bound transitions. Moreover, their ionization
energies are lower and they fall within frequencies with higher radiation flux.

\begin{figure*}
\begin{center}
\includegraphics[width=0.49\textwidth]{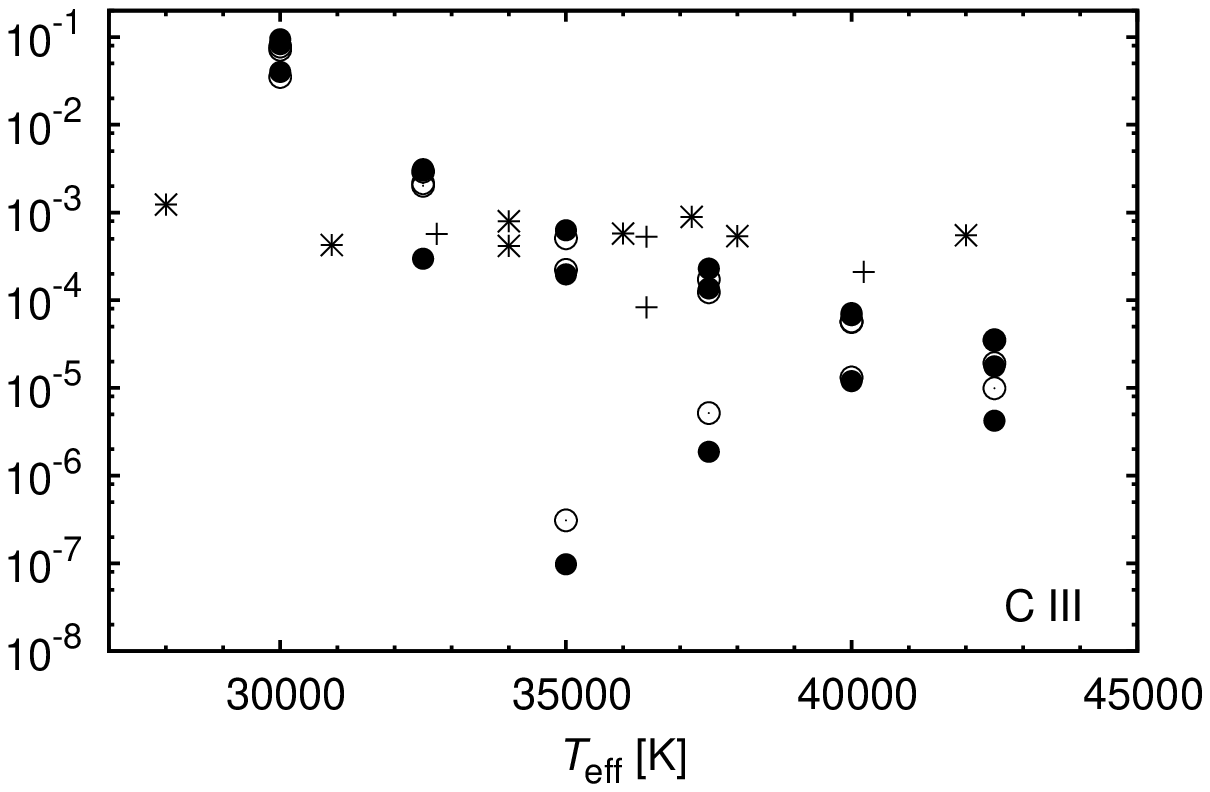}
\includegraphics[width=0.49\textwidth]{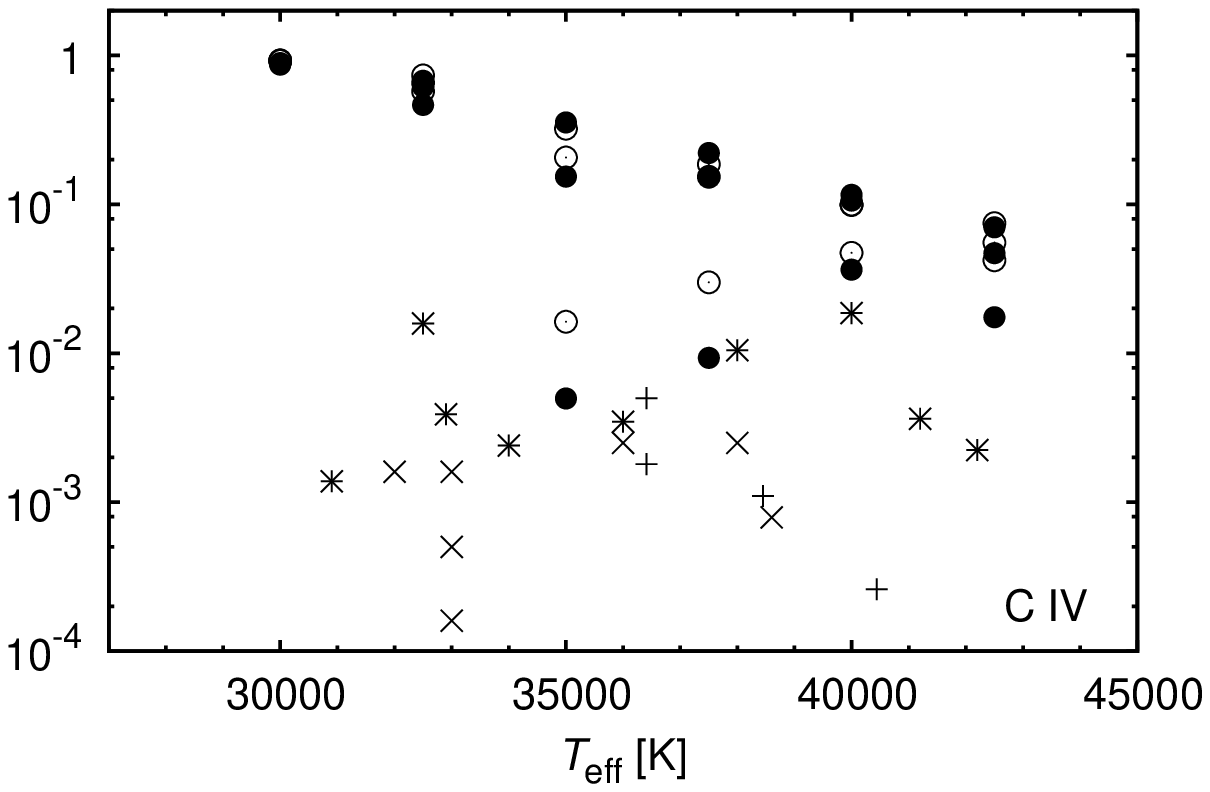}
\includegraphics[width=0.49\textwidth]{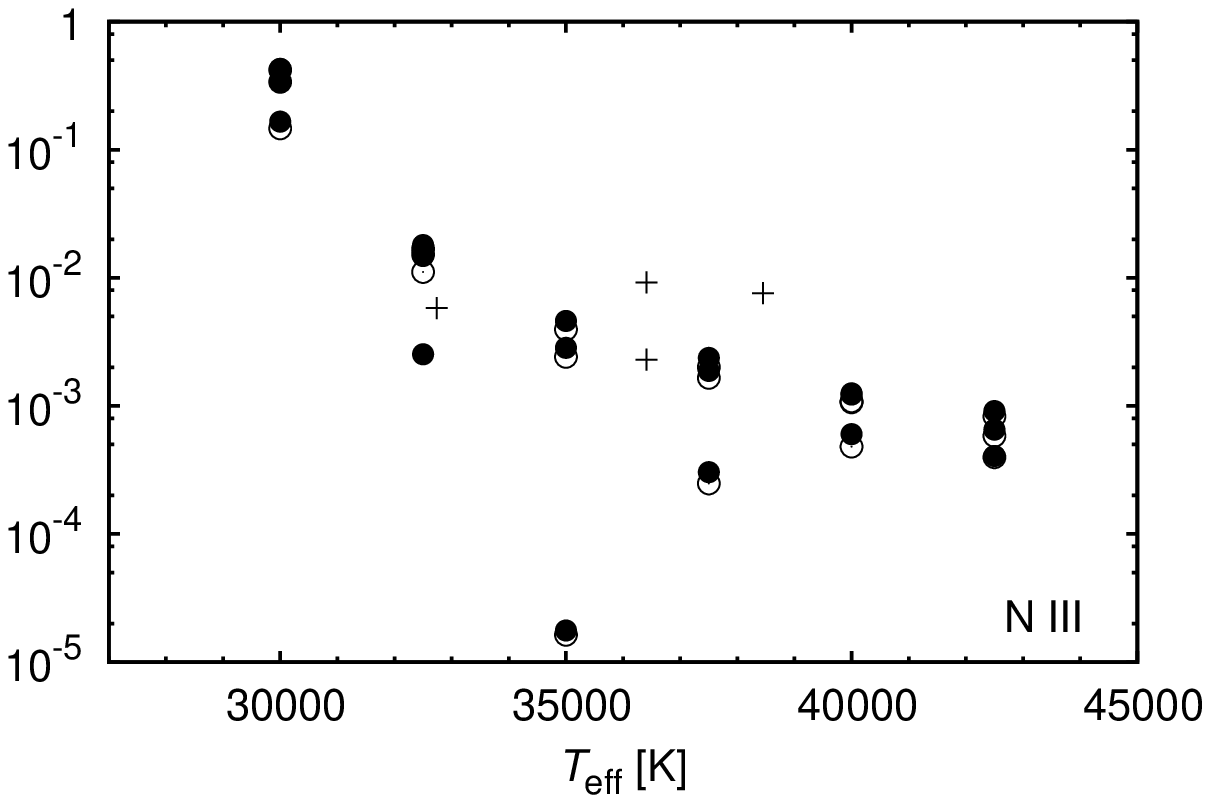}
\includegraphics[width=0.49\textwidth]{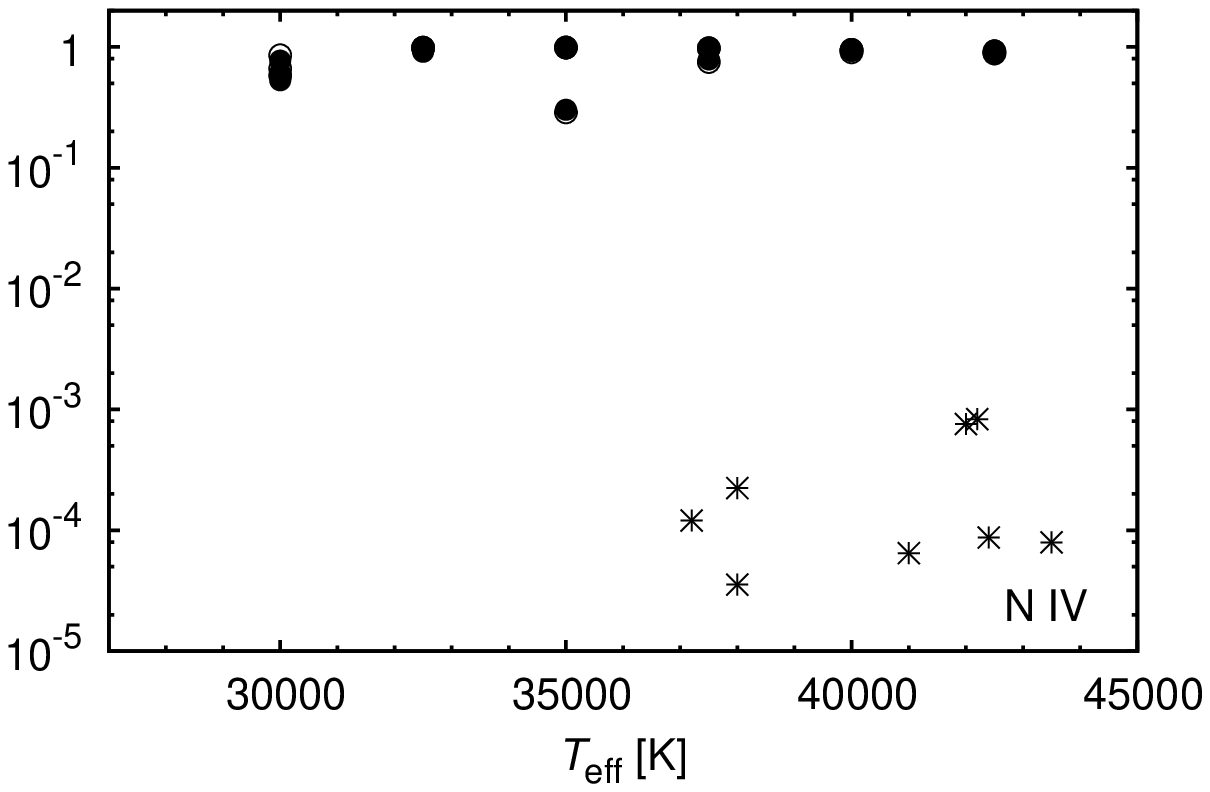}
\includegraphics[width=0.49\textwidth]{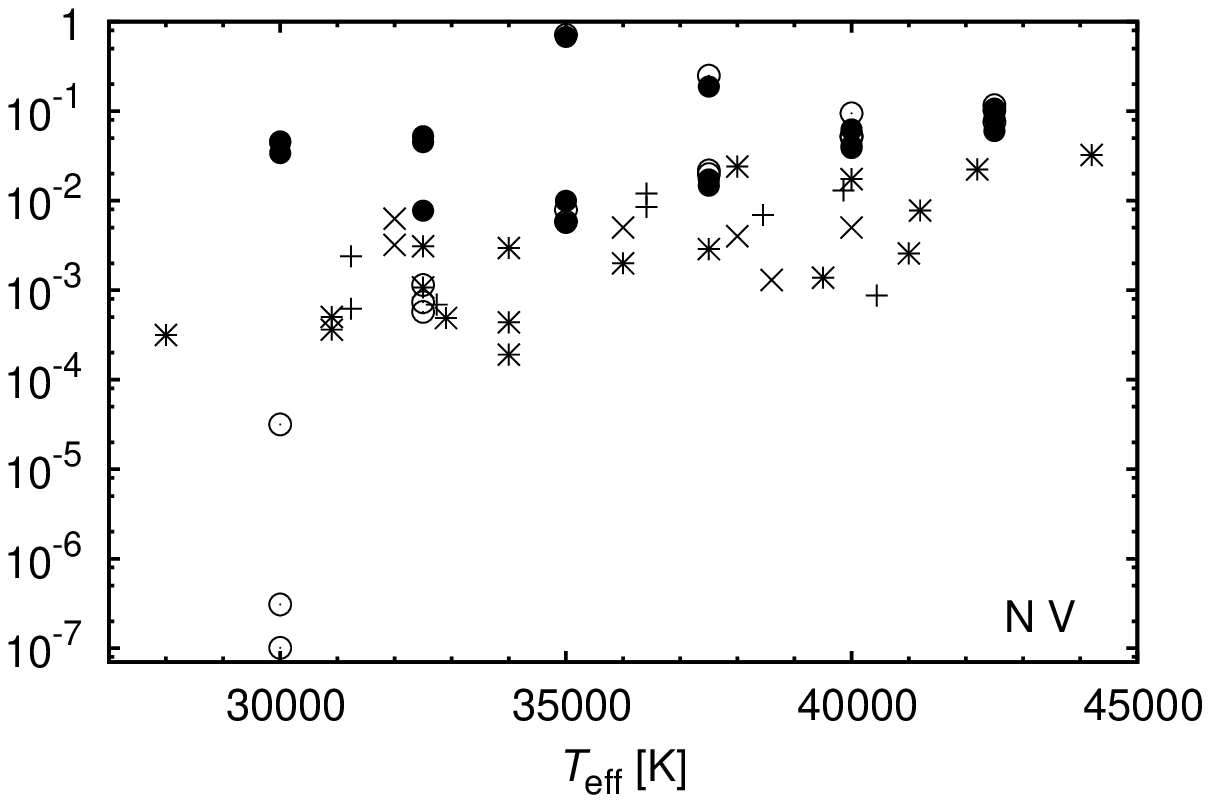}
\includegraphics[width=0.49\textwidth]{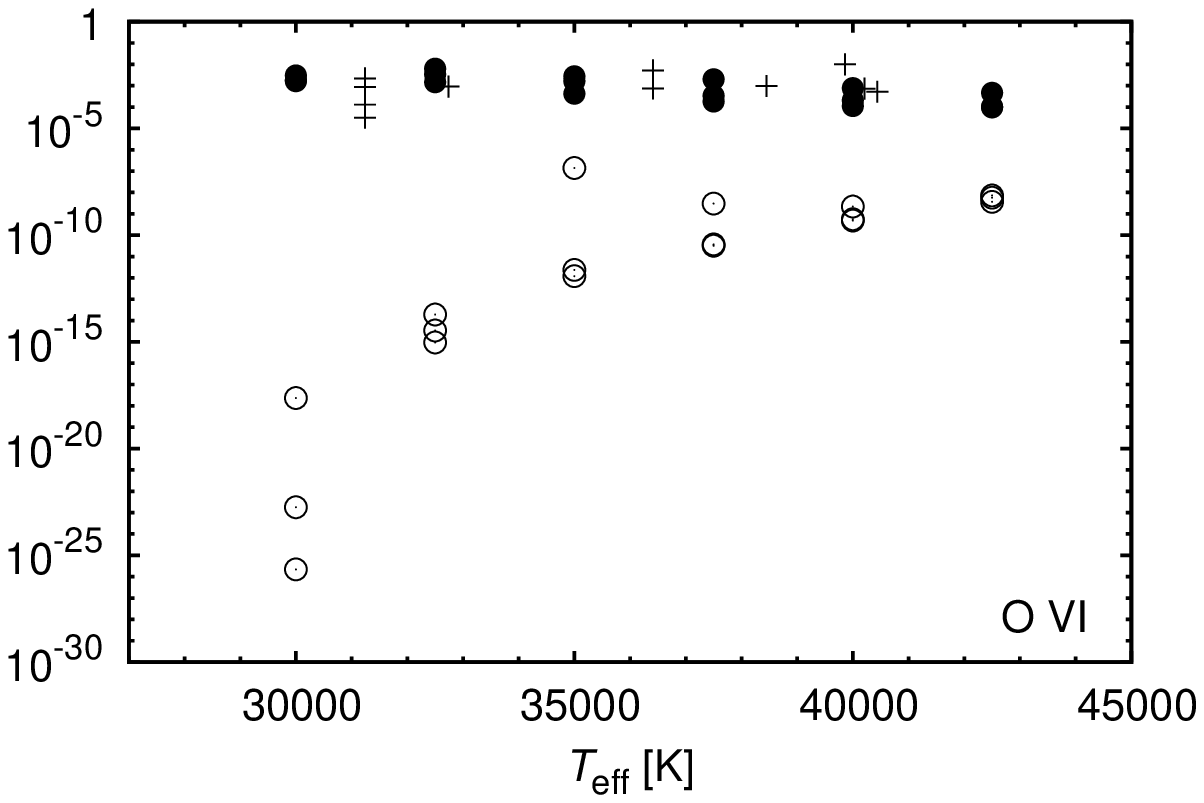}
\end{center}
\caption{Comparison of the ionization fractions calculated with additional X-ray
emission (full dots) and without X-ray emission (empty circles) at the point,
where the radial velocity is half of the terminal velocity, with observations
($+$ symbols: \citealt{masa2003}, $\times$ symbols: \citealt{howpri1989}, $*$
symbols: \citealt{lamion1999}) for CNO elements.}
\label{obrionicno}
\end{figure*}

\begin{figure*}
\begin{center}
\includegraphics[width=0.49\textwidth]{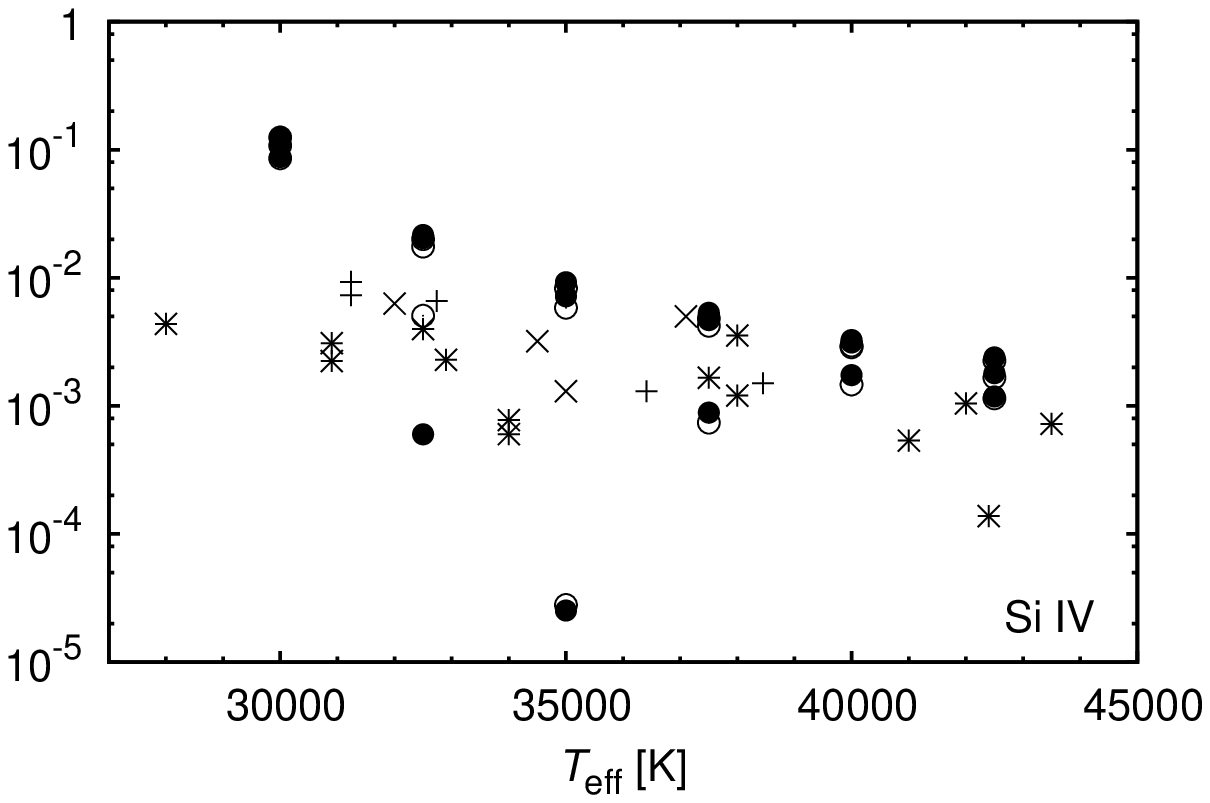}
\includegraphics[width=0.49\textwidth]{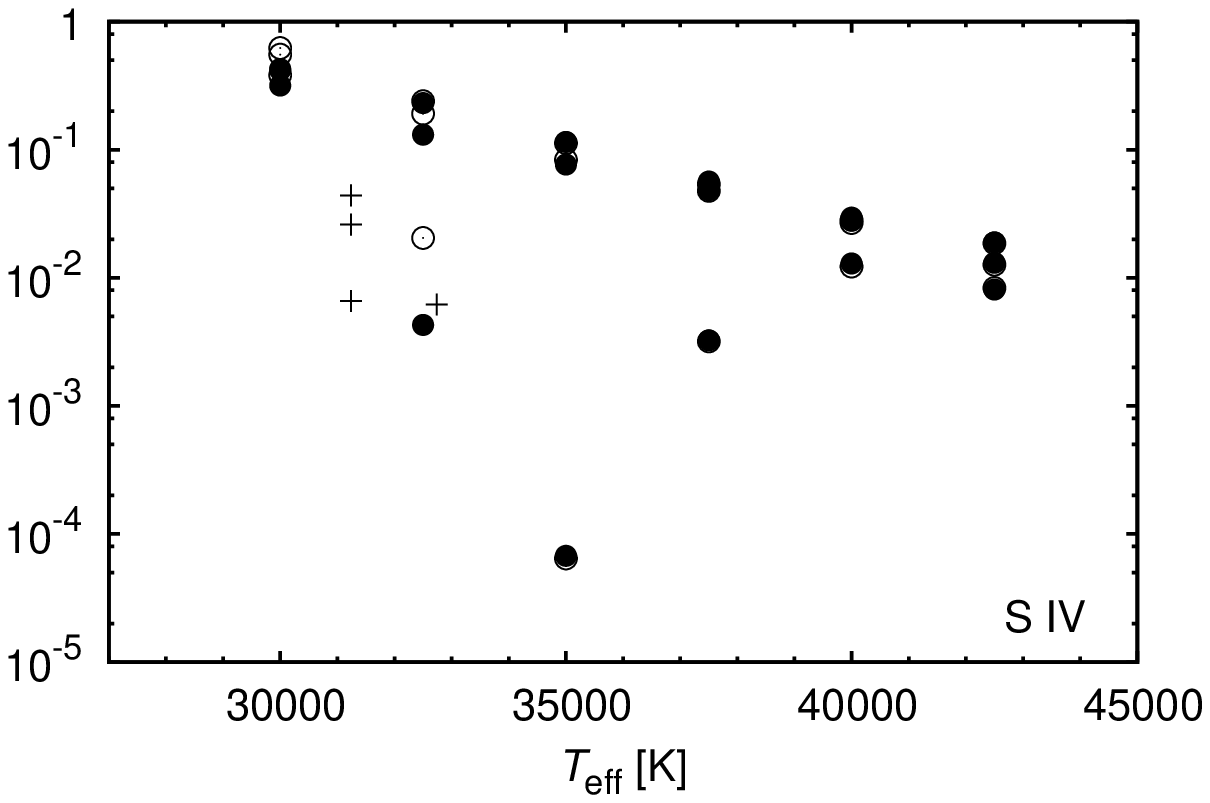}
\includegraphics[width=0.49\textwidth]{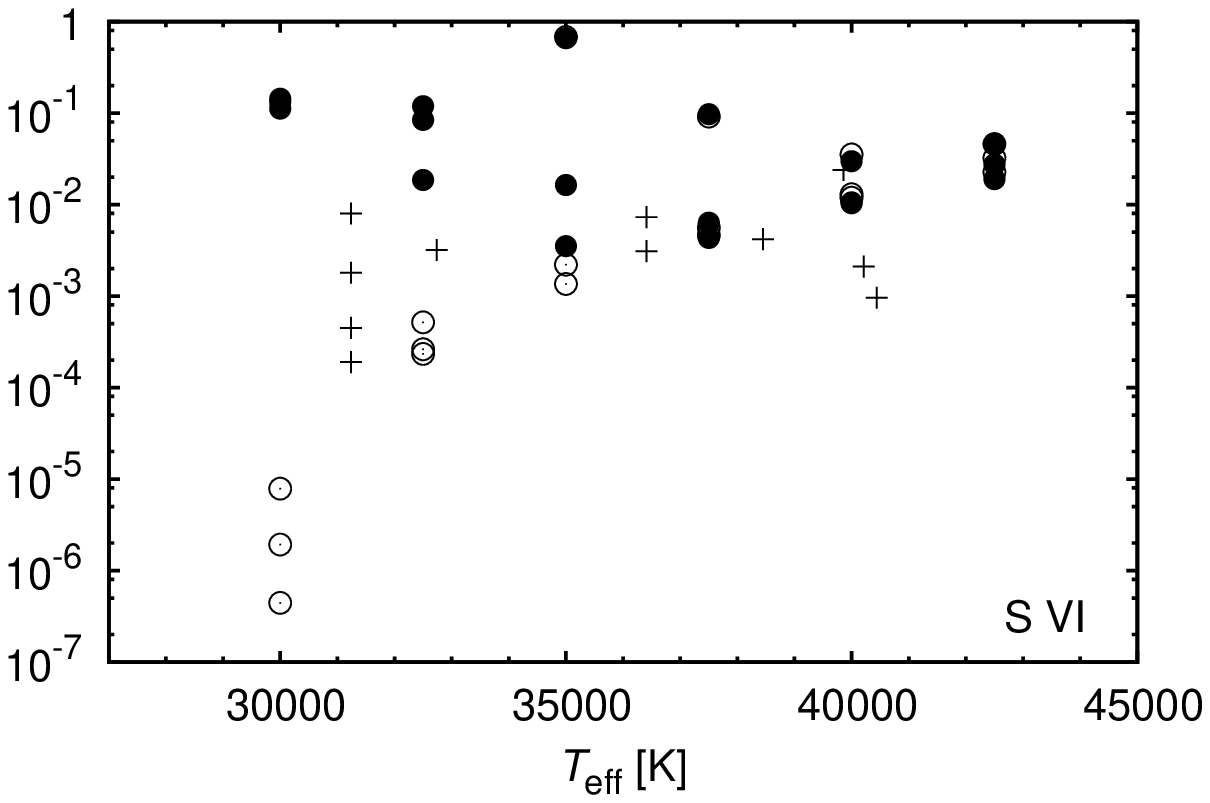}
\end{center}
\caption{Same as Fig.~\ref{obrionicno}, except for Si and S elements.}
\label{obrionisis}
\end{figure*}

The effect is even stronger for the \ion{C}{v} ion \citep{fosfor}. Consequently,
%Kr1: a
%
care has to be taken when making the conclusions about the existence of
additional ionizing radiation source just from the observations of \ion{C}{v} or
\ion{N}{v} lines. \cite{pasam} and \cite{mytri} showed that in many cases it is
possible to obtain sufficient abundance of \ion{N}{v} by proper treatment of the
kinetic equilibrium equations without any additional source of X-ray radiation.
Although it is generally true that
%Kr1:
an
additional ionization source shifts the
degree of ionization, due to the complexity of the processes involved in the
kinetic equilibrium equations, we can not claim that the presence of a
particular ion is caused exclusively by radiation at a chosen frequency. The
above example of \ion{C}{v} and \ion{N}{v} ionization fractions illustrates also
a potential importance of ionization by XUV radiation with frequencies larger
than
%Kr1:
the
\ion{He}{ii} ionization edge \citep[see also][]
%Kr1:{pakupa}.
{odecas,pakupa}.
%

%=======================================================================
\section{The case of single stars: influence of intrinsic X-ray radiation on the
ionization state}

We first discuss the influence of intrinsic X-rays created in the wind of single
stars. We assume that the X-ray generation is connected with 
%Kr1:
the
line-driving
instability \citep{maharaya,karluvhrad,abbvln,luciebila,oworyb,ovocnytep} which
steepens into highly supersonic shocks \citep{owcar,felpulpal,runow}. We use our
NLTE wind models with CMF line force, which consistently solve for the wind
hydrodynamical and thermodynamical structure.
%Kr1:
Our models include all relevant processes that affect the level populations,
i.e., radiative and collisional excitation and deexcitation, radiative and
collisional ionization and recombination, and the Auger processes. The
corresponding radiative ionization cross-sections are mostly taken from the 
Opacity and Iron Projects \citep{topt,zel0}.
%Ku2: consequently they
These cross-sections
include the
auto-ionization and dielectronic recombination.
In our models we include the fits
\citep[by][]{krfeloskuh} to the X-ray emission generated from the hydrodynamical
simulations of \citet{felpulpal}. The models and methods are described by
\citet{fosfor} in detail.

In Figs.~\ref{obrionicno} and \ref{obrionisis} we compare the predicted
ionization fractions of selected elements with observations. The comparison of
the ionization fraction calculated with and without X-ray emission shows that
X-rays do not significantly influence the ionization fraction of ions with low
ionization energies, such us \ion{C}{iv} or \ion{N}{v}, except for the lowest
effective temperatures and outer wind regions. Ionization fractions of ions with
higher ionization energies (e.g., \ion{O}{vi}) are affected by X-rays. The
\ion{N}{v} ionization fraction is influenced by X-rays only for cooler O stars
(and only in outer wind regions).

The ionization fractions derived from fitting of observed UV spectra with
theoretical ones (assuming wind mass-loss rates) in general agree with
ionization fractions predicted using hydrodynamic NLTE wind models. However,
there are some exceptions, for example the ionization fraction of \ion{N}{iv}
derived from observations is significantly lower than the predicted ionization
fraction (see 
%Kr2: Fig.~\ref{obrionisis}).
Fig.~\ref{obrionicno}).
A high ionization fraction of \ion{O}{vi}
(see Fig.~\ref{obrionicno}) derived from observations is a consequence of
X-rays, which can be nicely reproduced by NLTE models with additional X-ray
source.
%Kr1
%However, we note that the derived ionization fractions are also
%influenced by the local wind inhomogeneities \citep{sund,sur1,sur2}.
We also note that our smooth stationary wind models and the observational
%Ku2: diagnostic
diagnostics
of ionization fractions neglect inhomogeneities
which are present in the wind and which may be connected for example with
clumping \citep{sund,sur1,sur2}, or with effect of binary component
\citep[e.g.,][]{kapkar,vanlok}.

\begin{figure}
\begin{center}
\resizebox{0.9\hsize}{!}{\includegraphics{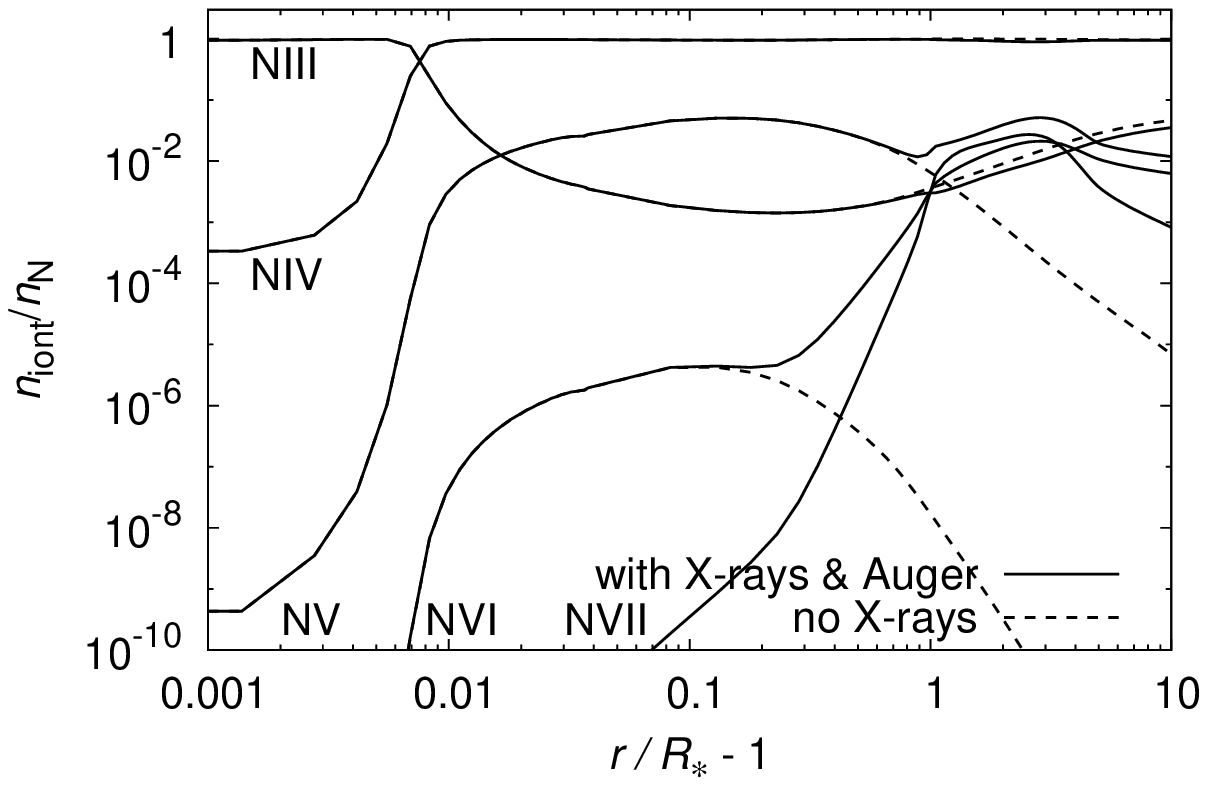}}
\resizebox{0.9\hsize}{!}{\includegraphics{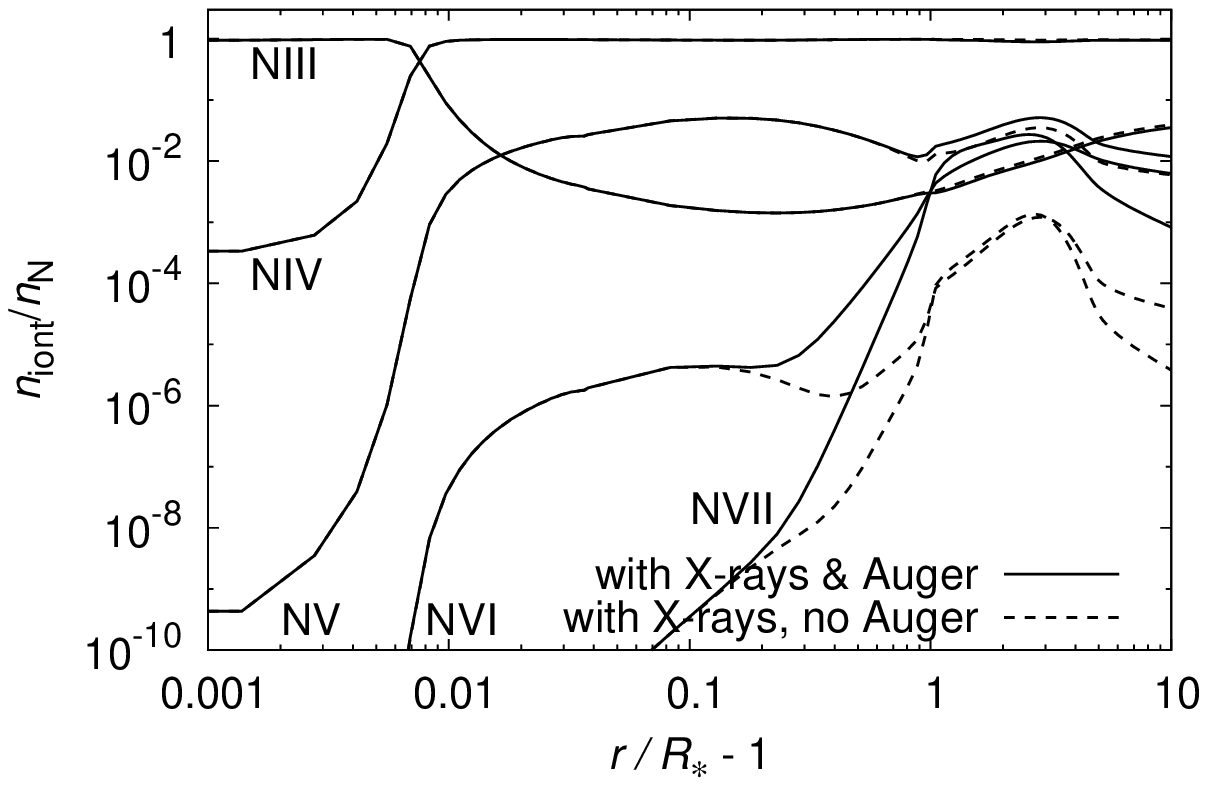}}
\end{center}
\caption{Variation of
%Kr1:
the
nitrogen ionization fractions with radius in the wind
model of
%Kr1:
a
supergiant with $T_\text{eff}=37\,500\,\text{K}$.
{\em Top panel:} Ionization fractions calculated with
%Kr1:
an
additional X-ray source (solid lines) and without the X-ray source (dashed
lines). {\em Bottom panel:} Ionization fractions calculated with additional
X-ray sources with inclusion of Auger processes (solid lines) and neglecting the
Auger processes (dashed lines).}
\label{375-1-ion}
\end{figure}

A further insight into the importance of individual ionization processes can be
obtained from the study of radial variations of ionization fractions. As an
example we plot the radial variations of
%Kr1:
the
nitrogen ionization fractions in the
supergiant wind model calculated by our code for
$T_\text{eff}=37\,500\,\text{K}$ in Fig.~\ref{375-1-ion}. Here we compare the
ionization structure of three different wind models: 
%Kr1:
a
model with
%Kr1:
an
additional X-ray
source and with Auger ionization, model with additional X-ray source but without
Auger ionization, and model without any additional X-ray source.

The region close to the stellar surface (for $r/R_\ast-1\lesssim0.1$) is opaque
for
%Kr1: the X-ray radiation.
X-rays.
Consequently, the ionization fractions of most abundant
ions (\ion{N}{iii}, \ion{N}{iv}) are not influenced by the additional sources of
X-rays close to the stellar surface and only minor higher ions (e.g.,
\ion{N}{vi} and \ion{N}{vii}) are affected by
%Kr1:
the
tiny amount of X-rays deeply
penetrating the wind (upper panel of Fig.~\ref{375-1-ion}). The wind regions
with larger radius (for $r/R_\ast-1\gtrsim0.1$) are less opaque for the
transmitted X-ray 
%Kr1: radiations
photons
and, consequently, they are more prone to change
their ionization state as a result of
%Kr1:
the
presence of X-rays. 

Individual nitrogen ions are influenced in different ways by X-rays. The
ionization fraction of \ion{N}{v} is increased as a result of the direct
ionization from \ion{N}{iv}, but is also affected by ionization to and
recombination from 
%Kr1: a
the
higher ion \ion{N}{vi}. The ionization fraction of
\ion{N}{vi} is increased by both Auger and direct ionization. \ion{N}{vii} is
generated due to the direct ionization of \ion{N}{vi} \citep{mytri}.

X-rays influence mostly minor ionization states in single hot stars,
consequently the wind mass-loss rate is not significantly influenced by
X-rays and the terminal velocity is typically only by 10--20\,\% percent
higher, especially in stars with weaker ionizing continua (with
$T_\text{eff}\lesssim35\,000\,\text{K}$, \citealt{mytri}).

%=======================================================================
\section{High-mass X-ray binaries: influence of X-rays on the radiative force}

\begin{figure}
\begin{center}
\resizebox{\hsize}{!}{\includegraphics{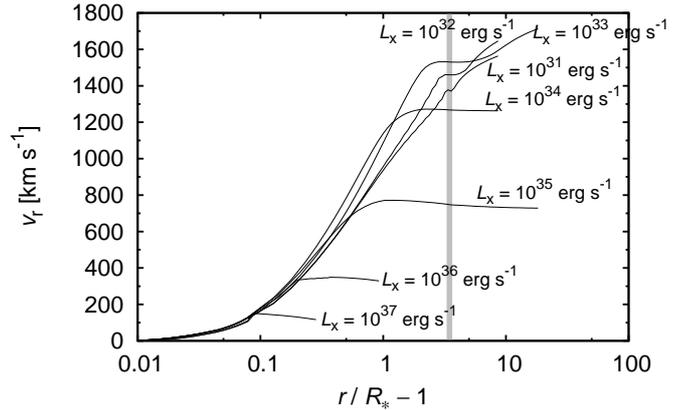}}
\end{center}
\caption{Radial variations of velocity in the model of supergiant with
$T_\text{eff}=30\,000\,\text{K}$ for different amount of external X-ray
irradiation (denoted in the graph). The irradiating source is located at
$r=100\,{R}_\odot$ (denoted by a vertical gray line).}
\label{300-1-r}
\end{figure}

The implications of X-ray irradiation may be more significant in X-ray binaries
due to their larger X-ray luminosities
\citep[e.g.,][]{viteal,hmxb40,dvoj18,zdarsky}. The influence of a strong X-ray
source, which is typical for binaries with
%Kr1:
an accreting
compact companion, is shown in
Fig.~\ref{300-1-r}. Here we plot the radial velocity in the supergiant wind with
$T_\text{eff}=30\,000\,\text{K}$ for different amounts of external X-ray
irradiation with
%Kr1:
an
irradiating source located at $r=100\,{R}_\odot$
\citep[see][for details]{dvojvit}. For low X-ray 
%Kr1: irradiation
%
luminosity,
$L_\text{X}\lesssim10^{31}\,\text{erg}\,\text{s}^{-1}$, X-rays do not
significantly influence the radiative force. For higher X-ray luminosities,
$L_\text{X}\gtrsim10^{32}\,\text{erg}\,\text{s}^{-1}$, X-rays influence the
dominant ionization states, which leads to a decrease of the radiative force
close to the X-ray source \citep[see, e.g.,][for observational
effects]{vanlok,viteal}. A typical kink in the velocity profile develops as a
result of weak radiative force \citep{feslo}. With increasing external source
X-ray luminosity the position of the kink moves towards the star and becomes
more pronounced. The X-ray ionization in the kink is so strong that the
radiative driving becomes inefficient and the wind stagnates. If this happens
for velocities lower than the escape speed, then the wind material may fall back
on the star
%Kr1:
\citep[c.f.,][]{posou}.
As soon as the kink reaches the critical point, where the mass loss
rate is determined, the X-rays lead to
%Kr1: the
%
wind inhibition. We define the
critical point as a point where the speed of the Abbott waves, the fastest waves
in the wind, is equal to the wind velocity \citep{abbvln,oworyb}.

\newrgbcolor{sarlatova}{0.79 0.37 0.37}
\newrgbcolor{plosar}{0.86 0.55 0.67}
\newrgbcolor{nejakazelena}{0.20 0.64 0.65}

\begin{figure}[t]
\centering
\resizebox{\hsize}{!}{\includegraphics{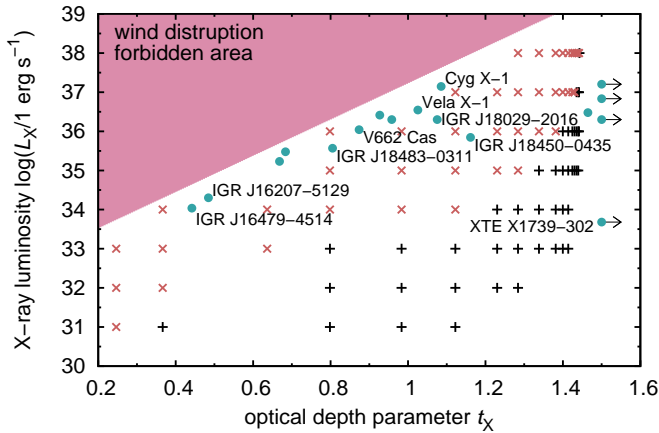}}
\caption{The diagram of X-ray luminosity (\lx) vs.~the optical depth parameter
($\x t$) displaying regions with different effects of the X-ray irradiation. The
graph is plotted for supergiant with $T_\text{eff}=30\,000\,\text{K}$.
Individual symbols denote positions of: models with negligible influence of
X-ray irradiation (black plus, $\boldsymbol+$), models where X-ray irradiation
leads to the decrease of the wind terminal velocity (red cross,
{\sarlatova$\boldsymbol\times$}), and non-degenerate components of HMXBs
(filled circles, {\nejakazelena\Large\raise
-2pt\hbox{\textbullet}}). Adopted from \citet{dvojvit}.}
\label{txobr}
\end{figure}

The influence of X-rays depends not only on the external X-ray luminosity, but
also on
%Kr1:
the
location of the X-ray source. Therefore, for stars with roughly the same
parameters (i.e., the same luminosity class and the temperature differences of
about 20\,\%) it is possible to construct a diagram of X-ray luminosity vs.~the
optical depth parameter (see Fig.~\ref{txobr}). The optical depth parameter
\citep{dvojvit}
\begin{equation}
\label{tx}
\x t=\frac{\dot M}{\varv_\infty}\zav{\frac{1}{R_*}-\frac{1}{D}}
\zav{\frac{10^3\,\text{km}\,\text{s}^{-1}\,1\,R_\odot}
{10^{-8}\,{M}_\odot\,\text{year}^{-1}}}
\end{equation}
characterizes the radial optical depth between the stellar surface and the X-ray
source. For a given star (with fixed mass-loss rate $\dot M$ and terminal
velocity $\varv_\infty$), the optical depth parameter $\x t$ in Eq.~\eqref{tx}
depends only on the X-ray source distance $D$. Winds with a negligible influence
of X-rays appear in the right bottom corner of Fig.~\ref{txobr}. Winds in which
the X-ray source causes a decrease in terminal velocity appear in a diagonal
strip 
%Kr1: of the 
in
Fig.~\ref{txobr}. A forbidden area, where the X-rays lead to wind
inhibition, is located in the upper left corner of Fig.~\ref{txobr}.

These results can be tested against the observed parameters of known high-mass
X-ray binaries (HMXBs). These stars are also included in Fig.~\ref{txobr}. The
positions of all HMXBs in  Fig.~\ref{txobr} lie outside the forbidden area in
agreement with our calculations. Moreover, many HMXBs appear close to the border
of 
%Kr1:
the
forbidden area. This may indicate that the winds of these stars are in a
self-regulated state. In such a case a higher X-ray luminosity is not possible,
because it would inhibit the flow \citep{velax1}. This keeps the star close to
the border of the forbidden area in Fig.~\ref{txobr}.

The influence of X-rays may be also estimated using the ionization parameter
\begin{equation}
\label{xi}
\xi(r)=\frac{1}{n d^2}\int L_\nu^\text{X}\text{e}^{-\tau_\nu(r)}\,\de\nu.
\end{equation}
Here $L_\nu^\text{X}$ is the X-ray luminosity per unit of frequency,
$\tau_\nu(r)$ is the optical depth between the X-ray source and a given point,
and the integration in Eq.~\eqref{xi} goes over the whole X-ray domain. The
exponential accounts for the absorption of X-rays in the wind. In the optically
thin limit Eq.~\eqref{xi} gives the well-known ionization parameter
$\xi\sim\lx/(n d^2)$ introduced by \citet[][see also \citealt{hacek}]{tenci},
where $\x L=\int L_\nu^\text{X}\,\de\nu$. Numerical analysis \citep{dvojvit}
shows
%Kr1:,
%
that for ionization parameters larger than about
$\xi=0.1-10\,\text{erg}\,\text{s}^{-1}\,\text{cm}$ the X-rays strongly inhibit
the radiative force. This may be modified in the presence of wind clumping,
which may weaken the effect of X-ray ionization because of increased
recombination inside clumps \citep{osfek}.

\citet{slanisko} propose
%Kr1:
an
alternative ionization parameter
\begin{equation}
\label{slanpar}
\eta(r)=\frac{n_\varphi(r)}{n_\text{e}(r)},
\end{equation}
where ${n_\varphi}$ is the number density of ionizing photons and $n_\text{e}$
is the electron number density. There is a relation between $\xi$ and $\eta$,
$\xi=4\pi c\eta\overline{h\nu}$, where $\overline{h\nu}$ is 
%Kr1: a
the
mean photon energy.
%Ku2: The 
Using the
%
%Kr1: advantage of the
%
ionization parameter $\eta$ given in
Eq.~\eqref{slanpar} is
%Kr1: that
%Ku2: somehow
%
advantageous, because
the photon number density directly enters the
kinetic equations. Therefore, the photon number density is in fact the basic
quantity and not the energy density. Moreover, $\eta$ is a nondimensional
parameter. Our numerical tests showed that the radiative force is inhibited by
X-rays for $\eta=10^{-4}-10^{-3}$.

%=======================================================================
\section{Implications for massive binaries with non-degenerate components}

In massive binaries with non-degenerate components the X-rays come from the
wind-wind collision \citep[e.g.,][]{usaci,kapitan,igor,skodik}. Consequently,
the X-rays can not be so strong to inhibit any of the winds of
%Kr1:
the
individual
components.

This is true in most X-ray binaries with non-degenerate components, in which the
parameters of individual components lie outside the forbidden area in
Fig.~\ref{txobr}. In many cases the parameters appear in the region with a
strong influence of X-rays on the radiative force and therefore also on the wind
velocity structure \citep{dvojvit}. These binaries may be in a self-regulated
state, but of a different type than we discussed in the case of HMXBs. In
binaries with non-degenerate components, an increase in X-ray luminosity \x L
causes a decrease in the wind velocity and therefore a decrease in \x L.
Similarly, the decrease in \x L leads to
%Kr1:
an
increase 
%Kr1: in
of
the terminal velocity,
which subsequently causes an increase in \x L \citep{pars}. This may be one of
the effects that lead to the observational finding that most non-degenerate
binaries are not stronger X-ray sources than corresponding single stars
\citep[e.g.,][]
%Kr4:{sane,igorkar}.
{oskikup,sane,igorkar}.

The parameters of some secondaries of massive binaries lie in the forbidden area
in Fig.~\ref{txobr}. This indicates that their wind is inhibited by the X-ray
emission. In some of them the X-ray emission may originate in the collision of
the primary wind with the secondary star surface \citep{dvojvit}.

%=======================================================================
\section{X-rays and the weak wind problem}

Theoretical models predict mass-loss rates that are in a fair agreement with
observations for stars with large mass-loss rates $\dot
M\gtrsim10^{-7}\,M_\odot\,\text{year}^{-1}$. However, for some stars with
mass-loss rates lower than this value the predicted mass-loss rates are by order
of magnitude higher than those derived from observations
\citep{bourak,martin,martclump}. This is 
%Kr1: so-called the
the so-called 
``weak wind problem".

The explanation of the ``weak wind problem" may lie in the low density of the
wind. As a result of the low wind density the shock cooling length may become
comparable with the hydrodynamical length scale, that is, the shocks change from
radiative to adiabatic \citep{owomix}. Consequently, there is a possibility that
once the wind of these stars is heated by the shocks it is not able to cool down
radiatively, and remains hot and therefore unaffected by the radiative
acceleration \citep{luciebila,martclump,cobecru,mytri,lucyjakomy}. Such shocks
occur well above the critical point \citep{owcar,felpulpal}, and consequently do
not influence the mass-loss rate. This means that the mass-loss rate of stars
with the ``weak wind problem" is roughly the same as the theory predicts. A high
temperature of the winds precludes reliable determination of the mass-loss rates
from UV lines. This conclusion may be supported by the observation of bow shocks
around stars with weak wind ($\zeta$ Oph and AE Aur), which require
significantly higher mass-loss rates than inferred from UV observations
\citep{gvar,grat}. 
%Kr1:
On the other hand, if the observed arc structures are
%Ku2: in fact
dust waves,
%Ku2: then
%
they may point to lower mass-loss rates \citep{ochsen}.

There is an alternative explanation of the ``weak wind problem". The X-rays of
stars with 
%Kr1: low density
low-density winds
may be so strong that they may lead to a significant
decrease of the radiative force and wind mass-loss rate \citep{nemajpravdu}.
This would mean that the mass-loss rate is indeed as low as deduced from
observations. However, \citet{dvojvit} argue that for realistic X-ray source
parameters the modification of the ionization equilibrium by X-rays is not
strong enough to significantly affect the mass-loss rate. This result is also
supported by \citet{oskibp}, who using their NLTE models  showed that the
observed level of X-ray emission in low-luminosity stars does not lead to such
%Kr1:
a
decrease of the radiative force and mass-loss rates that would explain the
problem with too weak wind line profiles.

\section{Conclusions}

Because the radiatively driven hot star winds are highly supersonic, they can
easily produce large amounts of X-rays. The X-rays influence the ionization
balance in the stellar wind. We discuss the influence of X-rays on the radiative
force and ionization equilibrium in hot star winds. The radiative force is not
affected by X-rays in single O stars. On the other hand, the X-rays in HMXBs are
so strong that they affect the radiative force, because highly ionized elements
are not able to accelerate the wind efficiently. This may even lead to
%Kr1: the
%
wind
inhibition. Therefore, there is a forbidden area of the binary parameters of
HMXBs, where the winds can not exist. Stellar wind of HMXBs may be in a
self-regulated state, where the X-rays control the hydrodynamical structure of
the wind.

\section*{Acknowledgments}

This research was supported by a grant GA\,\v{C}R  13-10589S.

%\clearpage

%\newcommand{\aap}{A\&A}
%\newcommand{\apj}{ApJ}

%\bibliography{rentgen}
%\bibliographystyle{model1-num-names.bst}
%\bibliographystyle{model1a-num-names.bst}
%\bibliographystyle{model2-names.bst}
%\bibliographystyle{model3-num-names.bst}
%\bibliographystyle{model4-names.bst}
%\bibliographystyle{model5-names.bst}
%\bibliographystyle{model6-num-names.bst}
%\bibliographystyle{elsarticle-num.bst}

\end{document}